\documentclass[journal]{IEEEtran}
\usepackage{cite}
\ifCLASSINFOpdf
\else
\fi

\usepackage[cmex10]{amsmath}
\usepackage{array}
\usepackage{mdwmath}
\usepackage{mdwtab}
\usepackage{graphicx}
\usepackage{euscript}
\usepackage{amsfonts}
\usepackage{bigstrut}
\usepackage{tabularx}
\usepackage{cite}

\usepackage{amsmath}
\usepackage{amsthm}
\usepackage{amssymb}
\usepackage{xcolor}

\usepackage{hyperref}

\usepackage{epstopdf} % When using eps figures

\usepackage{bbm}

\usepackage{enumitem}

\begin{document}

\title{Stochastic Geometry Methods for Modelling Automotive Radar Interference}
\author{Akram~Al-Hourani,~\IEEEmembership{Member,~IEEE}, Robin~J.~Evans,~\IEEEmembership{Life Fellow,~IEEE}, Sithamparanathan~Kandeepan,~\IEEEmembership{Senior Member,~IEEE}, Bill~Moran,~\IEEEmembership{Member,~IEEE}, and Hamid~Eltom~\IEEEmembership{Student Member,~IEEE}
\thanks{Manuscript received XX-March-2016 revised XX-XXXX-2016.}
\thanks{A. Hourani and R. J. Evans are with the Department of Electrical and Electronic Engineering at The University of Melbourne, Melbourne, Australia. E-mail: \{akram.hourani, robinje\}@unimelb.edu.au}
\thanks{S. Kandeepan, B. Moran and H. Eltom, are with the School of Engineering, RMIT University, Melbourne, Australia. E-mail: \{kandeepan.sithamparanathan, bill.moran, hamid.eltom\}@rmit.edu.au}
}

\markboth{Radar Interference}%
{Shell \MakeLowercase{\textit{et al.}}: Bare Demo of IEEEtran.cls for Journals}

\maketitle

\begin{abstract}
	As the use of automotive radar increases, performance limitations associated with radar-to-radar interference will become more significant. In this paper we employ tools from stochastic geometry to characterize the statistics of radar interference. Specifically, using two different models for vehicle spacial distributions, namely, a Poisson point process and a Bernoulli lattice process, we calculate for each case the interference statistics and obtain analytical expressions for the probability of successful range estimation. Our study shows that the regularity of the geometrical model appears to have limited effect on the interference statistics, and so it is possible to obtain tractable tight bounds for worst case performance. A technique is proposed for designing the duty cycle for random spectrum access which optimizes the total performance. This analytical framework is verified using  Monte-Carlo simulations.  
\end{abstract}


\IEEEpeerreviewmaketitle 

\section{Introduction}
\IEEEPARstart{A}{utomotive} radar is emerging as a key technology enabling intelligent and autonomous features in modern vehicles such as relieving drivers from monotonous tasks, reducing driver stress, and adding life-saving automatic interventions. Today, automotive radar is implemented in many high-end cars to enable essential safety and comfort features including adaptive cruise control and automatic emergency breaking systems where a vehicle can steeply decelerate without driver involvement to avoid a potential collision. The deployment of these features has thus far been limited to high-end vehicles because of the high cost of sensing technology such as automotive radar. However,  recent  advances in microelectronic technologies is creating the possibility of very low cost, high performance small radar sensors suitable for automotive applications \cite{6127923,Evans}. This advance will likely lead to wholesale  deployment of automotive radar in all classes of vehicles. %The radar-on-chip integrates millimetric radio components including antennas, A/D converters, mixers and amplifiers with the signal processing chip into a single small form factor board.

Given this upcoming vast deployment, it is anticipated that significant vehicle-to-vehicle radar interference will arise due to the shared spectrum use and the unavoidable geometry of road traffic situations such as on-coming traffic, intersections and turning. For example, a radar can be easily blinded or confused by vehicles travelling in the opposite direction resulting in degraded performance in radar detection ability that might coincide with a split-second critical road situation. Similarly, backward looking radars can interfere with forward looking radars for vehicles  travelling in the same direction. Interference can also arise due to  multiple reflections in dense traffic and in built-up areas. Intersections are a further obvious source of radar-to-radar interference. The interference in all of these cases is largely caused by the use of shared spectrum and the inherent lack of coordination between radars resulting from the lack of centralized control and resource allocation. There exist, of course, many tools to handle radar interference including clever waveform design, fast adaptive antenna methods particularly nulling, polarization switching, various signal processing methodologies, and many more. We are currently examining a number of these approaces in an endeavour to address the automotive radar interference problem.

In this paper we study certain aspects of the stochastic behaviour of automotive radar interference by modelling road vehicles as a spatial point process. We utilise tools from stochastic geometry to formulate an analytic framework characterising the arising interference in terms of its cumulative distribution function and mean value based on a given vehicle density in a road segment. This framework is further employed to understand the average performance of automotive radar in terms of \emph{ranging success probability}, that is the probability of reliably detecting a target given a certain set of operating conditions. We develop explicit formulae that tightly characterize the lower bound performance and provide insight into the different dynamics influencing performance. Furthermore we introduce a new metric for measuring the overall performance of radars operating within a given spectral bandwidth, we call this metric the \emph{spatial success probability} and we utilize it to develop a method for finding the optimum duty cycle for any specific radar to randomly accesses spectrum resources. 

\subsection{Contribution}
The main contributions of this paper are
\begin{itemize}
%\item a novel approach to model road vehicles based on stochastic geometry.
\item an analytic framework for estimating the level of  interference experienced by a radar under certain specified scenarios
\item an analytic framework for calculating the expected signal-to-interference ratio and hence the expected performance of automotive radar.
\item an optimization methodology to calculate the optimum value of the random spectrum access duty cycle.
\item an intuition showing the convergence of Bernoulli lattice towards Poisson point process.
\item closed-form expressions providing insight into various system dynamics that contribute to determining automotive radar performance under certain specified situations.
\end{itemize}

\subsection{Paper Organization}
The remainder of this paper is organized as follows. Section \ref{Sec_Litrature} provides relevant background and an overview of recent results on the application of stochastic geometric methods for modelling vehicle locations. In Section \ref{Sec_Network_Model} we develop appropriate geometric and propagation  models. In Section \ref{Sec_Interference} we present an analytical approach to characterise radar interference for certain scenarios, leading to Section \ref{Sec_Performance} where we determine the expected radar performance statistics and introduce a performance optimization methodology. A simulation procedure is explained in Section \ref{Sec_Simulation}. Finally we provide concluding remarks  in Section \ref{Sec_Conclution}. The main notation and symbols are summarized in Table \ref{Table_Notations} for convenience.

\section{Background and Related Work}\label{Sec_Litrature}
The expected vast global  market penetration of automotive radar technology has required both international and local regulatory authorities to work in conjunction with the automotive industry to develop appropriate and harmonized standards. It is anticipated that by 2030 the penetration of automotive radars will reach around 65\% in Europe and  50\% in US as described in the International Telecommunication Union (ITU) document \cite{ITU_UWB}. In another recommendation document \cite{ITU_AutomotiveRadar}, ITU classifies automotive radar into two main categories according to their ranging capabilities and safety requirements:
\begin{itemize}
  \item \textbf{Category 1}: Designed for long distances up to 250 m, serving the adaptive cruise control and collision avoidance systems. This category is the main focus of the analytic work presented in our paper.
  \item \textbf{Category 2}: Designed for short and medium distances up to 50-100 m depending on the application, utilized for lane change assistance and rear traffic crossing alert,
\end{itemize}

The  bandwidth requirement of the long range category is planned to be 1 GHz, with maximum allowed Equivalent Isotropic Radiated Power (EIRP) of 55 dBm. The medium/short range category has less power allowance, and a wider spectrum bandwidth to support higher range resolution for close targets, based on the standard resolution and bandwidth relation \cite{6544299}, $\Delta R = \frac{c}{2B}$, where $\Delta R$ is the range resolution, $c$ is the speed of light and $B$ is the used bandwidth.

There is now a growing activity aimed at understanding and addressing the problem of mutual interference arising from overlapping automotive radar signals. Many of these attempts have been initiated by industry such as the EU project MOSARIM \cite{6450766} which investigated automotive radars interference by conducing experimental road measurements and by conducting complex ray-tracing simulations. This particular  project also explored some possible interference mitigation techniques. An important conclusion from this project suggests that (particularly for LFM radar waveforms)  interfering radars are unlikely to cause ghost targets but rather they will create noise-like combined interference. 

In this paper we focus on noise-like interference and consider the signal-to-interference ratio as the key factor in determining radar performance. Ghost targets due to interference were studied analytically in \cite{4106078} and were observed when two perfectly-identical radars are utilized with identical waveforms. However, the same paper found that it is more likely that two radars will cause a noise-like interference under practical scenarios. Exploiting this observation, the works in \cite{6544299} and \cite{6525508} suggest a practical approach for randomizing chirp sweep frequency in order to guarantee noise-like interference, thus aiming to reduce the false alarm probability caused by ghost targets. Random frequency step (RFS) radar is also suggested in the literature to mitigate radar-to-radar interference such as the work in \cite{4137843}, suggesting that RFS would also suppress range ambiguity and enhance covert detection. Practical system algorithms to efficiently implement RFS in automotive applications has been filed by our team in the patents \cite{morelande2013method,evans2014apparatus} suggesting reduced interference when utilizing this scheme.

Further analytic attempts to investigate automotive radar interference can be found in \cite{7045843} which studies the desired-to-undesired signal power ratio in ultra wideband automotive radar, also in \cite{5760761} and \cite{7117925} utilizing Frequency Modulation Continuous Wave (FMCW) as a modulation scheme. Simulation approaches can be found in \cite{7024937} and \cite{7117928}, mainly based on ray tracing with scenario specific simulation environments. To summarize, our understanding of the available literature on automotive radar interference, we list the following points:
\begin{itemize}
  \item The majority of the literature is based on simulation and empirical approaches.
  \item Some analytic approaches investigate the interference in simple scenarios consisting of two vehicles.
  \item Simulation approaches investigate interference based on complex ray-tracing and stochastic environments.
  \item Most of the literature uses \emph{modulation-specific} simulations and analysis, namely FMCW, and pulse radar.
\end{itemize}

Stochastic geometry can be used to characterise the randomness in the positions of vehicles. Much recent literature has exploited the tractability facilitated by stochastic geometry tools to analyse the performance of wireless networks where it captures the spatial randomness of wireless network elements such as the location of base stations and the location of users. This approach differs significantly from the often used case-specific simulations such as the standard hexagonal simulation models set by 3GPP\footnote{3GPP is the third generation partnership project, mainly leading the efforts of cellular networks standardization.} to compare the performance of different vendors. Recent papers such as \cite{Andrews_Capacity} and \cite{Al-Hourani2015} characterise the interference in cellular networks where the signal-to-interference ratio (SIR) is estimated and the average network throughput is deduced. Stochastic geometry is also used for analysing ad-hoc and sensor networks \cite{Cite_Baccelli_Aloha,Akram_Globecom2}. The work in \cite{Blaszczyszyn_1D} and \cite{Blaszczyszyn_Conference} applies stochastic geometry in vehicular ad-hoc networks (VANET) to determine the average transmission success rate assuming that road vehicles are spatially distributed according to a linear Poisson point process (PPP). A single dimension Poisson point process is also adopted in \cite{ElSawy2015}, however the paper uses a multi-lane linear PPP in order to enhance the model accuracy for wider highways. The work in \cite{Tong_Haenggi} disregards the effect of the road width for modelling the IEEE802.11p standard, and uses a modified version of the Mat\'{e}rn hard-core point process to capture the effect of media access control based on a PPP vehicle distribution.

To the best of our knowledge, the work presented in this paper is the first to model the stochastic behaviour of automotive radar interference based on analytic tools from stochastic geometry. Our approach can be applied to a wide range of waveform techniques since it deals with the media as a limited resource pool (temporal, spectral or code). Moreover, it is the first to provide an estimate of the radar ranging success probability based in the interference statistics in closed-form expression. The optimization of spectrum access duty-cycle is also a novel approach that does not appear to have been previously addressed.

\section{System Model}\label{Sec_Network_Model}
In this section we construct the system model that emulates the geometric layout of vehicles on a road and the associated radio propagation environment. Without loss of generality, we consider a vehicle located at the origin and call it the \emph{typical vehicle}, and assume that its statistical behaviour is typical of all other vehicles on the road. Furthermore, we consider a temporal snapshot of the road traffic during which the vehicles can be considered as stationary, where having another snapshot should preserve the geometric statistics of the traffic. We note here that these geometrical statistics are indeed not constant in the long run, however they are of a slow kinetic nature and can be safely thought of as static for a given segment of the road over a reasonable observation period.

\subsection{Geometrical Model}
A modern vehicle could be fitted with more than a one radar \cite{6127923}, typically a long range radar (LRR) for distances of 10-250 m, and several  short/medium range radars (MRR/SRR) for distances of 1-100m and 0.15-30m respectively. The LRR radar is mounted on the front of the vehicle providing vital information for the automatic cruise control and collision avoidance/mitigation systems. The MRR/SRR are mounted on the sides and back of the vehicle. This paper focusses on the LRR.

A simplified layout of the interfering LRR radars is illustrated in Fig. \ref{Fig_Layout} showing a typical vehicle with the potential interfering radars travelling in the opposite direction. Taking into consideration the defined narrow antenna pattern and ignoring sidelobes, the interfering vehicles are the ones located beyond a minimum distance $\delta_o$, expressed as:
\begin{equation} \label{Eq_delta_o}
\delta_o = \frac{L_n}{\tan \frac{\theta}{2}},
\end{equation}
where $\theta$ is the antenna beamwidth, $L_n$ is the distance between the lane of the typical vehicle and the $n^{\text{th}}$ opposing lane, where multiple opposing lanes can exist. %This depends on the default antenna pattern but it serves as a suitable approximation for the purpose of interference analysis.

%Taking the fact distance $L_n$ is much smaller than the length of the road segment, in addition to the fact that the antenna beamwidth of the frontal radar is usually limited to $15^\circ$ \cite{ACMA_Radar}.% we can approximate the vehicle layout in Fig. \ref{Fig_Layout} into a one-dimensional geometry as indicated in Fig. \ref{Fig_Model}.

A simplified \emph{macroscopic} behaviour of the traffic \cite{6701151} considers the density of vehicles  as a function of both the location and time $\Lambda(x,t)$ measured in vehicles per unit length. However, by looking at a certain temporal snapshot and by considering a homogeneous average vehicle density, we can model the density as a constant value $\lambda$ in a certain road segment.

We capture the randomness in the locations of vehicles in two extreme geometrical distributions (point processes). In the first case we assume complete irregularity in the locations of vehicles with no correlation between these locations. This case is modelled by a Poisson point process (PPP) with intensity $\lambda$. The second extreme  occurs when vehicles are located on a deterministic lattice layout, i.e. periodic locations in space, separated by a constant distance $\delta$. In this layout vehicle locations are considered to be completely regular. In practical scenarios we expect that the actual distribution of vehicles will reside between these two geometrical extremes. We shall show in this paper that under practical road parameters, radar system performance is tightly bounded by these two extreme cases.

Based on the preceding assumptions, we depict a simplified geometrical layout in Fig. \ref{Fig_Model} showing both point processes and indicating the typical vehicle location and  interfering vehicle locations. We now consider the two extreme point processes further:
\begin{figure}[!t]
\centering
\includegraphics[width=\linewidth]{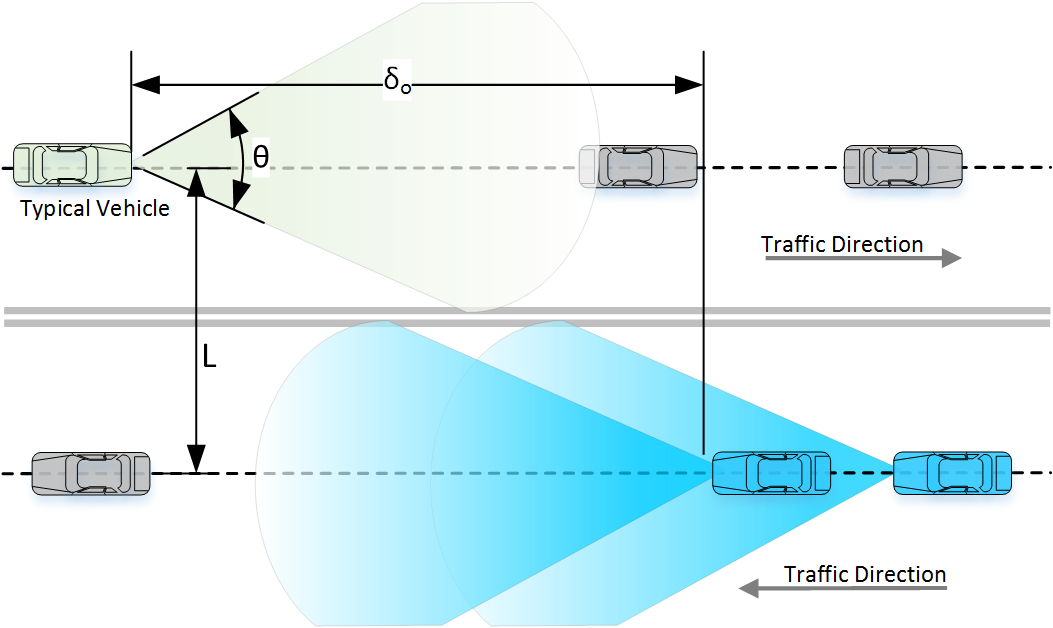}
\caption{A simplified geometrical layout indicating the typical vehicle under study and interfering vehicles (in blue).}
\label{Fig_Layout}
\end{figure}
%Following are the adopted two point processes explained in details:

\subsubsection{Poisson Point Process Model}
One extreme of the geometrical distribution is achieved when vehicle locations on a certain lane are completely independent of each other. This case resembles a unidimensional Poisson point process (PPP) \cite{Haenggi_Matern} in $\mathbb{R}^1$ with a homogeneous linear intensity $\lambda$ measured in vehicles per unit length. We denote this set of vehicles as $\Phi_\mathrm{PPP}$. Utilising a PPP allows a tractable analysis to be developed using Campbell theorems \cite{Book_Stoyan}. To model the effect of medium access, thus to capture the effect of concurrently transmitting vehicles, we apply \emph{random thinning} on the opposing vehicles set $\Phi_\mathrm{PPP}$ with a retention probability equal to the probability of resource collision given as $\xi$, representing the potential that an opposing vehicle is concurrently utilising the same resources as the typical vehicle, thus $\xi$ can be thought of as the duty-cycle of the random spectrum access. Applying a random marking on $\Phi_\mathrm{PPP}$ we can describe the interferers set as:
\begin{equation}
\Theta_{\mathrm{PPP}} = \{ x : x \in \Phi_{\mathrm{PPP}} , \mathcal{M}(x)=1\},
\end{equation}
where the mark $\mathcal{M}(x)$ is defined as:
\begin{equation}\label{Eq_Marking}
\mathcal{M}(x)=  \left\{
\begin{array}{lr}
	0                   & : x \leq\delta_o \\
	\boldsymbol{B}(\xi) &   : x >\delta_o,
\end{array}
\right.
\end{equation}
where vehicles closer than $\delta_o$ are marked as non-interfering, and $\boldsymbol{B}(\xi)$ is a Bernoulli random variable with selection probability $\xi$, where random variables in this paper are denoted in {\bf bold}.  %that is by referring to the model depicted in Fig. \ref{Fig_Layout} where the interfering vehicles are the one travelling in the opposite direction and are located beyond the distance $\delta_o$, and utilising the same radio resources as of the typical vehicle (the same frequency, time, or code) they will cause interference to the typical vehicle located at the origin. 

Note that the points of $\Theta_{\mathrm{PPP}}$ are located in the domain $(\delta_o,\infty)$ and have a reduced intensity of $\lambda_\Theta = \xi \lambda$ , this model is depicted in the upper part of Fig. \ref{Fig_Model}. Note that the independent thinning of a PPP yields another PPP \cite{Stoyan_Book}.

\subsubsection{Lattice Model}
\begin{figure}[!t]
\centering
\includegraphics[width=\linewidth]{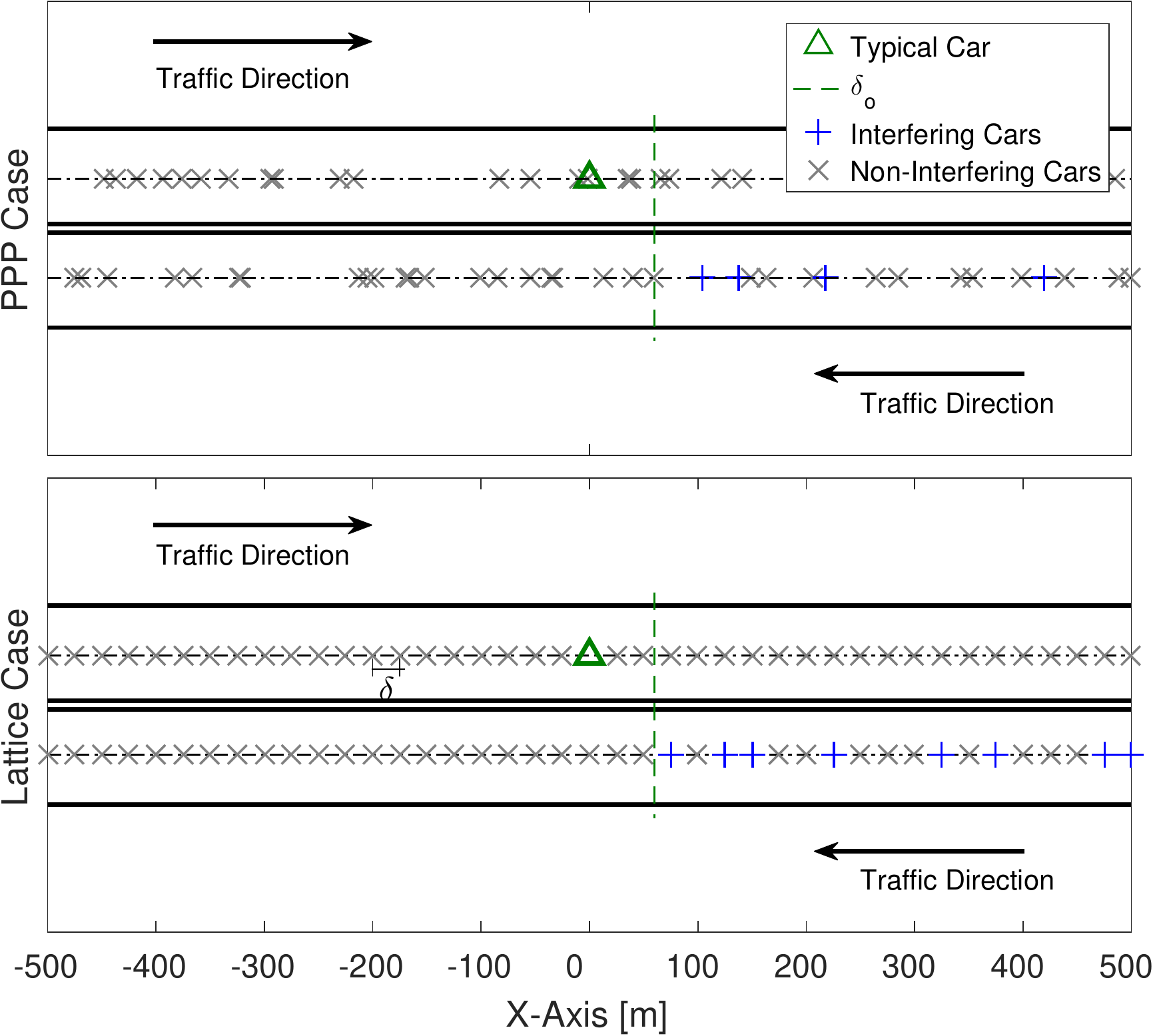}
\caption{The proposed geometrical models: (up) a linear Poisson point process, and (down) a regular  lattice.}
\label{Fig_Model}
\end{figure}

In this model we assume that vehicles are distributed according to a deterministic (regular) one-dimensional lattice, where vehicles can only take discrete locations with predefined spacing distance $\delta$ unit length. Thus having a linear density of $\lambda=\frac{1}{\delta}$. Although vehicle locations in the same lane are deterministic, they would exhibit no correlation to the typical vehicle in the opposing lane, thus a uniformly distributed random variable (RV) is randomly translating the entire lattice in a linear manner. Accordingly, we may express the set of approaching vehicles as:
\begin{equation}
\Phi_{\mathrm{TL}} = \{ (x+\boldsymbol{\mathcal{U}} )\delta + \delta_o ~ : ~ x \in \mathbb{Z}\},
\end{equation}
where $\mathbb{Z}$ is the integers set, and $\boldsymbol{\mathcal{U}}$ is a standard uniformly distributed random variable in the range $[0,1]$. Noting that $\boldsymbol{\mathcal{U}}$ is single random variable (not a vector) that captures the randomness in the grid translation with respect to the typical vehicle, where all approaching vehicles are translated with an equal value of $\boldsymbol{\mathcal{U}} \delta$. Following (\ref{Eq_Marking}) we can \emph{mark} the subset of interfering vehicles:
\begin{equation}
\Theta_{\mathrm{BL}} = \{ x : x \in \Phi_{\mathrm{TL}}, \mathcal{M}(x)=1\},
\end{equation}
where the subscript BL means \emph{Bernoulli Lattice}. As depicted in the lower part of Fig. \ref{Fig_Model}, where interfering vehicles are indicated with a blue (+) sign.

\subsection{Channel Model}
The electromagnetic energy transmitted by the radar travels in a propagation environment that  adds extra losses on top of the natural wavefront expansion. These losses are caused by atmospheric attenuation and absorption. Instead of using the simple inverse square law for estimating the power decay, we model the RF propagation using a general decay exponent $\alpha$. % and a stochastic channel fading factor $\boldsymbol{g_o}$ having unity mean.

We utilise the concept of signal-to-interference-plus-noise ratio (SINR) to evaluate the performance of the radar system. Understanding that the arbitrary interference generated by other vehicles can be perceived as white noise in the receiver. Thus, we first characterise the \emph{ranging signal} power i.e. the signal travelling form the radar towards the target and bouncing back, then we characterise the \emph{interfering signals} originated from all interfering vehicles, as further elaborated below:

\subsubsection{The Ranging Signal}
\begin{figure}[!t]
\centering
\includegraphics[width=\linewidth]{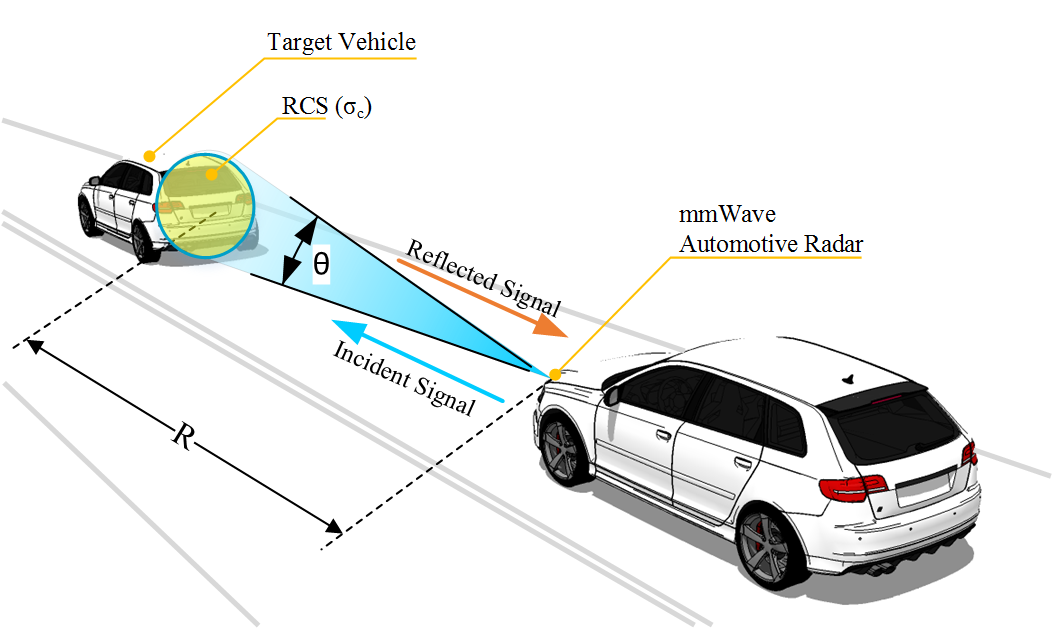}
\caption{Illustration of the frontal long range automotive millimetre-wave radar.}
\label{Fig_Radar_Equation}
\end{figure}

The raging signal transmitted by the radar and reflected from the target(s) is well-characterised in the literature to follow what is called the \emph{radar equation} \cite{Book_Radar}, however in order to improve the resilience of our model we consider an inverse $\alpha$-law instead of the common inverse square-law in order to incorporate the atmospheric absorption in the millimetre-wave spectrum \cite{Akram_ICC_2014}. Absorption in the 76-77 GHz can be estimated, according to ITU Recommendation ITU-R P.676-4 \cite{Cite_ATM_Absorbtion}, to vary between 0.09 - 0.15 dB \cite{ACMA_Radar} for practical radar detection rage up to 300 m. Thus, the path-loss exponent $\alpha$ can be seen to slightly exceed 2 in most  practical cases.% In addition, we consider a statistical fading process in the channel denoted as a random variable $\boldsymbol{g_o}$, where we assume a reciprocal channel sustaining its characteristics within the sensing turn-around time. This randomness is caused by multipath propagation and by unwanted echoes from buildings, vegetation, other vehicles and weather effects as rain. If unwanted reflections arrive within the integration (acquisition) time of the required signal, then they will contribute to the fading process. Accordingly the assumed $\boldsymbol{g_o}$ can include several underlying processes that corresponds to \emph{radar clutter} \cite{Book_Radar} and multipath fading.
The resulting modified radar equation in the millimeter-wave can then be written as,
%\begin{equation}\label{Eq_Modified_Radar_Equation}
%S = \underbrace{\boldsymbol{g_o} \frac{P_o G_t}{4\pi R^\alpha}}_{\text{Incident signal}} \times \underbrace{{\boldsymbol{g_o} \frac{\sigma_c}{4\pi R^\alpha}A_e}}_{\text{Reflected signal}} =\boldsymbol{g_o}^2 \gamma_1 \gamma_2 P_o R^{-2\alpha}
%\end{equation}
\begin{equation}\label{Eq_Modified_Radar_Equation}
S = \underbrace{\frac{P_o G_t}{4\pi R^\alpha}}_{\text{Incident signal}} \times \underbrace{{\frac{\sigma_c}{4\pi R^\alpha}A_e}}_{\text{Reflected signal}} =\gamma_1 \gamma_2 P_o R^{-2\alpha}
\end{equation}
which models the back reflected signal strength, where $P_o$ is the radar transmit power, $R$ is the target range i.e. the distance to the target, $G_t, A_e$ are the antenna gain and the effective area respectively, and $\sigma_c$ is the \emph{radar cross-section} area (RCS) of the target. The parameters $\gamma_1$ and $\gamma_2$ are given as,
\begingroup
\allowdisplaybreaks
\begin{align} \label{Eq_Gammas}
	\gamma_1 &= \frac{G_t Ae}{4\pi} = G_t^2\left(\frac{c}{4\pi f}\right)^2, \\
	~\text{and}~ \gamma_2 &= \frac{\sigma_c}{4\pi},
\end{align}
\endgroup
where $f$ is the operating frequency. Note that in the standard radar equation \cite{koks2014create}, the inverse square law is used i.e. $\alpha=2$. We illustrate the dynamics affecting the radar signal in Fig. \ref{Fig_Radar_Equation}.

\subsubsection{The Interfering Signals}
We utilise similar channel conditions for the interfering signals, by following an $\alpha$-law for the mean signal decay and by incorporating stochastic channel fading and clutter effects in an i.i.d. random vector $\boldsymbol{g_x}$, where $x$ refers to the particular interferer. Accordingly the resultant interference power at the radar of interest caused by an interferer $x$, is given by,
\begin{equation}
I_x = \gamma_1 P_o \boldsymbol{g_x} ||x||^{-\alpha},
\end{equation}
where $||x||$ refers to the Euclidean distance measured from the origin to $x$. The combined interference is addressed in the following section.

\section{Radar Interference} \label{Sec_Interference}
Assuming that radars are mounted on the front of vehicles and are fitted with directive antennas, the interference is mainly caused by approaching vehicles driving in the opposite direction to the typical vehicle, and the aggregated interference at the typical vehicle can be written as,
\begin{align}\label{Eq_8}
\boldsymbol{I}   &= \sum_{x \in \Theta} P_x  = \sum_{x \in \Theta} \gamma_1 P_o \boldsymbol{g_x}||x||^{-\alpha},
\end{align}
where $\Theta \in \{\Theta_\mathrm{PPP},\Theta_\mathrm{BL}\}$ is the point process describing the interferers. In the following subsections we describe the statistics of interference in terms of its mean, standard deviation and cumulative distribution function.

\subsection{Interference Mean}
In order to find the mean (expectation) value of the interference, we apply the Campbell theorem \cite{Haenggi_Book} to calculate the sum over a PPP:
\begin{align}\label{Eq_9}
\overline{I_\mathrm{PPP}}=\mathbb{E}\left[\boldsymbol{I}\right]      &= \mathbb{E}_g \mathbb{E}_{\Theta_\mathrm{PPP}} \left[\sum_{x \in \Theta_\mathrm{PPP}} \boldsymbol{g_x} \gamma_1  P_o ||x||^{-\alpha} \right] \nonumber \\
                     &\stackrel{(a)}{=} \mathbb{E}_g[g] \int_{\delta_o}^\infty \lambda_I \gamma_1 P_o u^{-\alpha} \mathrm{d} r \nonumber \\
                     &= \mathbb \int_{\delta_o}^\infty \lambda_I \gamma_1 P_o u^{-\alpha} \mathrm{d} r \nonumber \\
\end{align}
where $\mathbb{E}_g$ is the expectation over the channel stochastic process, $\mathbb{E}_{\Theta_\mathrm{PPP}}$ is the geometric expectation over all possible realizations of the interferers locations, and $\lambda_I = \xi \lambda$ is the effective intensity (density) of the interferers. The variable $u=\sqrt{r^2+L_n^2}$ represents the distance between the typical vehicle and the interferers in the n$^\mathrm{th}$ opposing lane. Step (a) follows the assumption that individual propagation channels have an i.i.d distribution which is independent from the the geometrical point process. The final step assumes that the average channel gain is normalized to unity i.e $\mathbb{E}[g]=1$. Evaluating the integral yields,
\begin{align}
\overline{I_\mathrm{PPP}}=\xi \lambda \gamma_1 P_o L_n^{-\alpha }
&\left[\frac{2 \sqrt{\pi } L_n \Gamma \left(\frac{\alpha +3}{2}\right) } {\left(\alpha ^2-1\right) \Gamma \left(\frac{\alpha }{2}\right) } \right. \nonumber \\
& \quad\quad\left. - \delta_o \, _2F_1\left(\frac{1}{2},\frac{\alpha }{2};\frac{3}{2};-\frac{\delta_o^2}{L_n^2}\right)
\right]
\end{align}
% Equation reverified 20160126 with Mathematica
where $_2F_1\left(.,.;.;.\right)$ is the hypergeometric function and $\Gamma(.)$ is the Gamma function. If we neglect the lane spacing when compared to the longitudinal distance $r$. We can obtain,
\begin{equation} \label{Eq_11}
\overline{I_\mathrm{PPP}}|_{L_n\to 0} = \frac{\xi \lambda  \gamma_1 P_o}{(\alpha-1)\delta_o^{\alpha-1}} \text{ ,\quad for } \alpha > 1
\end{equation}
Based on the above relation, we note that the interference is linearly proportional to the effective interferer density $\xi\lambda$. Also we note the strong effect of  $\delta_o$, take for example the default case when $\alpha \approx 2$ then interference is inversely proportional to $\delta_o$, thus it is clear that a narrower antenna beamwidth will increase $\delta_o$ and reduce the interference.

The interference arising from a translated Bernoulli lattice of interferers $\Theta_\mathrm{BL}$ can be calculated starting from (\ref{Eq_8}),
\begingroup\makeatletter\def\f@size{7}\check@mathfonts
\begin{align}
\overline{I_\mathrm{BL}}&=\mathbb{E}\left[\boldsymbol{I}\right] \nonumber \\
     &= \mathbb{E}_g \mathbb{E}_{\mathcal{U}}\mathbb{E}_{\Theta_\mathrm{BL}} \left[\sum_{x \in \Theta_\mathrm{BL}} \boldsymbol{g_x} \gamma_1 P_o ||x||^{-\alpha} \right] \nonumber \\
                     &= \mathbb{E}_g\mathbb{E}_{\mathcal{U}} \mathbb{E}_{\Theta_\mathrm{BL}} \left[ \sum_{m=0}^{\infty}  \boldsymbol{g}_m\boldsymbol{B}_m(\xi) \gamma_1 P_o \left[L_n^2 + (\delta_o +(m+\boldsymbol{\mathcal{U}})\delta)^2\right]^{-\frac{\alpha}{2}}\right]        \nonumber \\
                     &=\xi  \gamma_1 P_o  \sum_{m=0}^{\infty} \mathbb{E}_{\mathcal{U}} \left[\left[L_n^2 + (\delta_o +(m+\boldsymbol{\mathcal{U}})\delta)^2\right]^{-\frac{\alpha}{2}}\right]
\end{align}
\endgroup
where $\mathbb{E}_g[g] = 1$ as it has been used in (\ref{Eq_9}), emphasising that $\boldsymbol{\mathcal{U}}$ is a single random variable and not a vector. Similar to the PPP case, if we ignore the lane distance when compared to the longitudinal span of the road, i.e. $L_n\to 0$, then the expectation over $\boldsymbol{\mathcal{U}}$ can be  evaluated and the interference will take the simpler form of, 
\begin{align}\label{Eq_13}
&\overline{I_\mathrm{BL}} |_{L_n\to 0} \stackrel{(a)}{=}  \nonumber \\
&\xi  \gamma_1 P_o \delta^{-\alpha}  \sum_{m=0}^{\infty} \delta \frac{\left(A+m\right)^{1-\alpha}-(A+m+1)^{1-\alpha}}{\alpha-1} \nonumber \\
&= \frac{\xi  \gamma_1 P_o \delta^{-2\alpha}}{\alpha-1} \left[\zeta\left(\alpha-1,A \right)-\zeta\left(\alpha-1,A+1 \right)\right] \nonumber \\
&\stackrel{(b)}{=} \frac{\xi  \gamma_1 P_o \delta^{-2\alpha}}{\alpha-1} A^{1-\alpha}=\frac{\xi \lambda  \gamma_1 P_o}{(\alpha-1)\delta_o^{\alpha-1}} \nonumber \\
&=\overline{I_\mathrm{PPP}}|_{L_n\to 0} ,
\end{align}
where $A$ is given by $A=\frac{\delta_o}{\delta}$, and $\zeta(s,a) = \sum_{m=0}^{\infty} (m+a)^{-s}$ is the Hurwitz zeta function. Step (a) results from applying the expectation over $\boldsymbol{\mathcal{U}}$, while step (b) follows from the fact that $\zeta(s,a)=\zeta(s,a+1)+a^{-s}$ as per equation (25.11.3) in \cite{olver2010nist}. The result in (\ref{Eq_13}) indicates that the average interference from a translated Bernoulli lattice is exactly equal to its counterpart on the linear Poisson point process, thus the regularity of points (vehicles) does not have an effect on the average interference.

\begin{figure}[!t]
	\centering
	\includegraphics[width=\linewidth]{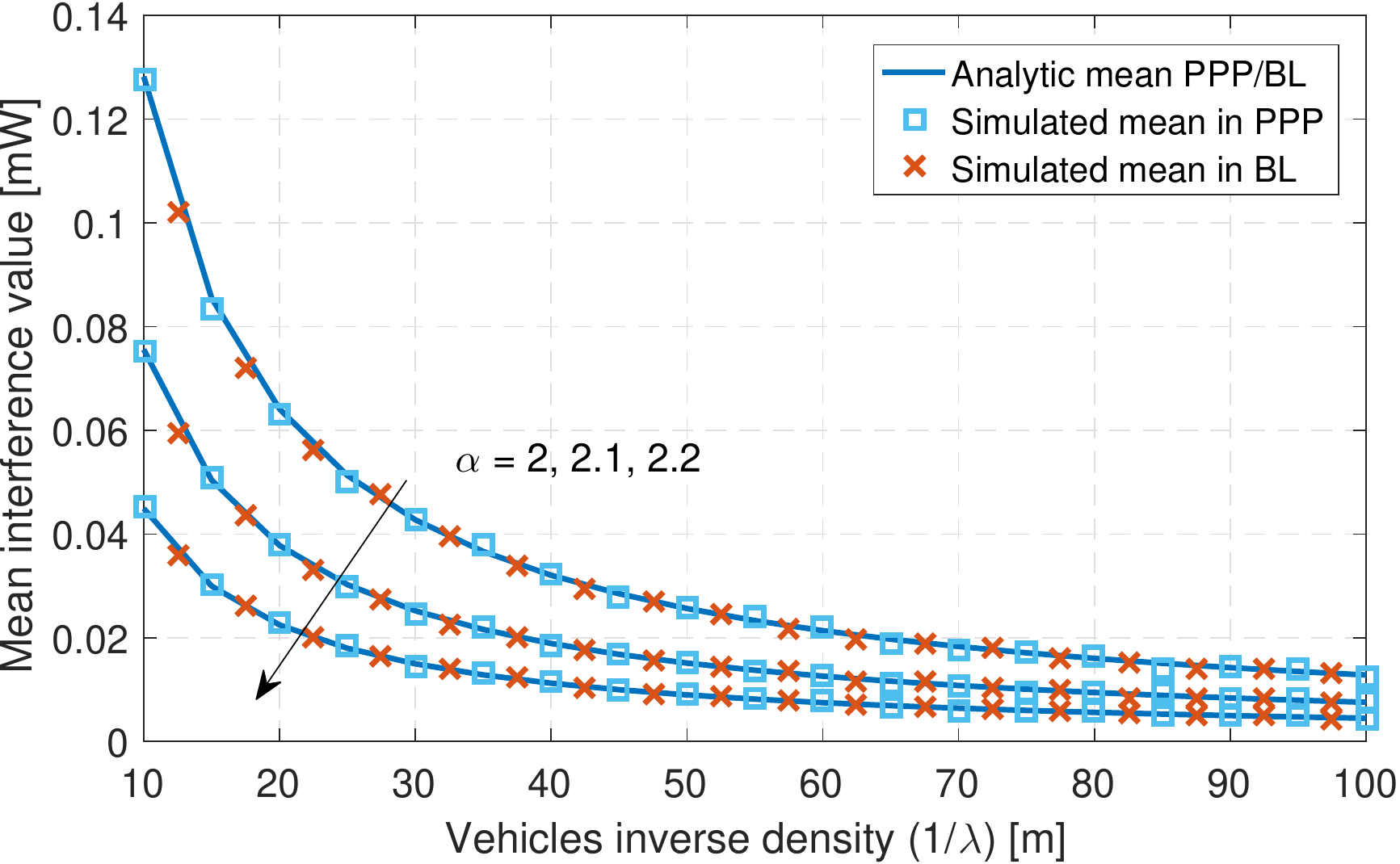}
	\caption{A comparison between the mean interference of the two point processes models; PPP in (\ref{Eq_11}) and BL in (\ref{Eq_13}) for $L_n=0$, and numerical parameters in Table. \ref{Table_Notations}.}
	\label{Fig_I_Mean}
\end{figure}

We plot in Fig. \ref{Fig_I_Mean} a comparison between the interference arising from a simulated PPP field from one side and by a BL from another side, the plot indicates the matching behaviour of the two models together, also matching the closed form expression in (\ref{Eq_13}). The numerical parameters are listed in Table \ref{Table_Notations}. Later in Sec \ref{Sec_Int_Dist} we shall provide an intuitive explanation of the convergence of BL process to PPP under certain conditions.

\subsection{Interference Distribution} \label{Sec_Int_Dist}
\begin{figure}[!t]
	\centering
	\includegraphics[width=\linewidth]{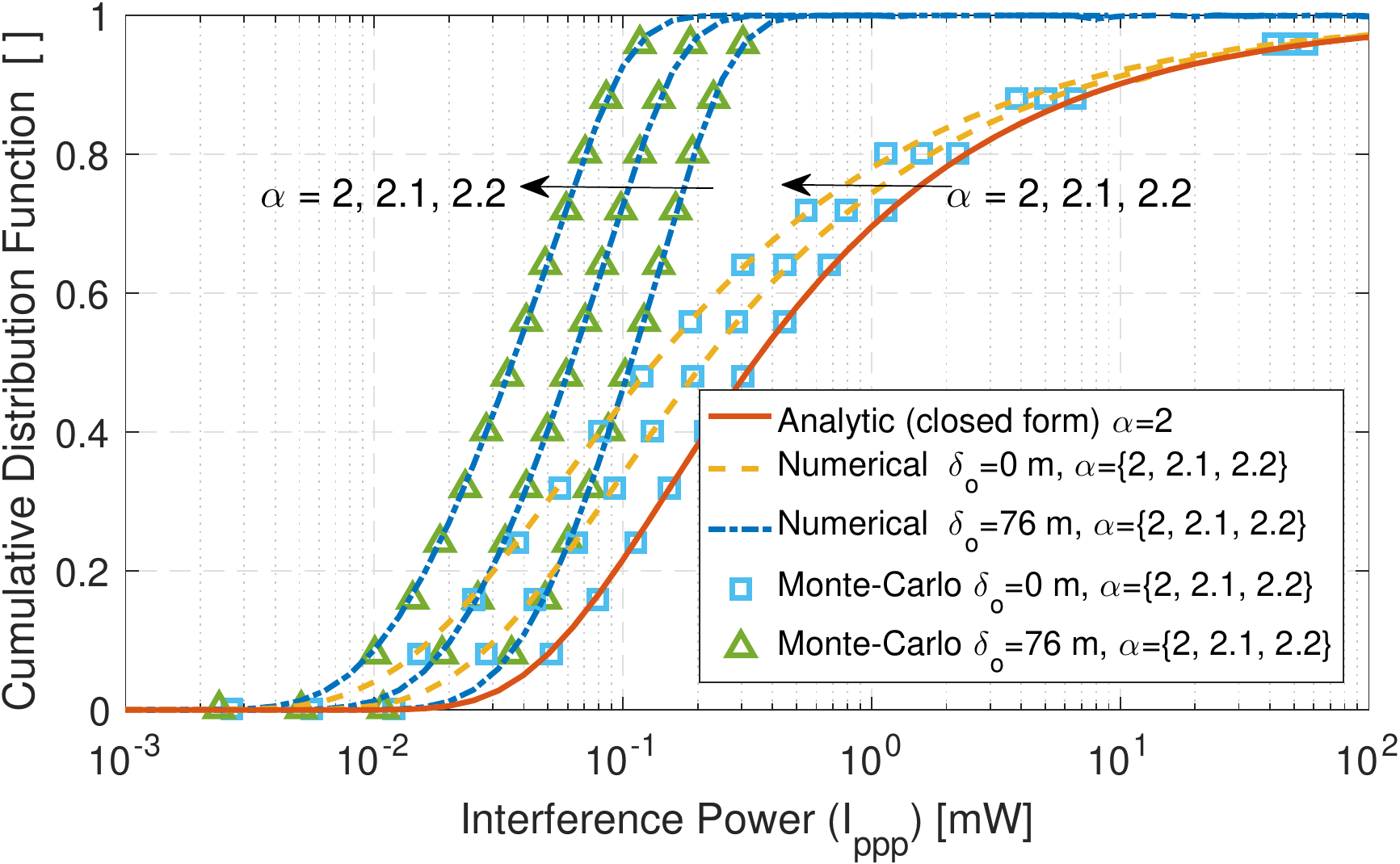}
	\caption{The CDF of the radar interference arising from the geometrical PPP model, comparing Monte-Carlo results with the analytic and numerical plots receptively obtained from equation (\ref{Eq_22}) and by using the inversion theorem in (\ref{Eq_Inversion_Theorem}). Parameters as per Table \ref{Table_Notations}, and $\frac{1}{\lambda_I}=100$ m.}
	\label{Fig_I_CDF_PPP}
\end{figure}

In order to get a deeper insight into the statistical behaviour of the interference, we obtain its characteristic function (CF) and then formulate its cumulative distribution function (CDF). We start from the definition of the CF of a random variable $\boldsymbol{x}$ as $\varphi_{\boldsymbol{x}}(\omega) = \mathbb{E}\left[ e^{\jmath \omega \boldsymbol{x}} \right]$. Accordingly, the characteristic function of the interference in case of PPP model can be expressed as,
\begingroup\allowdisplaybreaks
\begin{align}\label{Eq_Derivation_1}
&\varphi_{\boldsymbol{I}_{\text{PPP}}}(\omega) =  \mathbb{E}  \left[ e^{\jmath\omega\boldsymbol{I}} \right] \nonumber \\
&=\mathbb{E}_{\boldsymbol{g}} \mathbb{E}_{\Theta_\mathrm{PPP}} \left[ \exp\left(\jmath\omega\sum_{\Theta_\mathrm{PPP}} \boldsymbol{g_x} \gamma_1 P_o ||x||^{-\alpha} \right) \right] \nonumber \\
&= \mathbb{E}_{\Theta_\mathrm{PPP}}  \left[ \prod_{\Theta_\mathrm{PPP}}\mathbb{E}_{\boldsymbol{g}}  \exp\left(\jmath\omega \boldsymbol{g_x} \gamma_1 P_o ||x||^{-\alpha} \right) \right].
\end{align}
\endgroup
The last step follows from the fact that the channel random variable is independent of the geometrical stochastic process. Now we can apply the following probability generating functional of the homogeneous PPP \cite{Haenggi_Book},
\begin{equation}\label{Eq_PGFL_PPP}
E\left[ \prod_{\mathcal{I}} f(x) \right] = \exp\left(   - \int_{\mathcal{R}} [1-f(x)] \Lambda(\mathrm{d}x)  \right),
\end{equation}
where $\Lambda(\mathrm{d}x)$ the intensity measure on the infinitesimal volume $\mathrm{d}x$. In our case $\Lambda(\mathrm{d}x)=\lambda_I\mathrm{d}u$, where $u$ is the integration variable. This integration is performed over the region  of interest $\mathcal{R}$, where in our case it represents the active interferers region $\mathcal{R} \in [\delta_o,\infty]$. Thus, using (\ref{Eq_PGFL_PPP}) we can continue to simplify (\ref{Eq_Derivation_1}) where we drop the $x$ notation in $\boldsymbol{g}_x$ since it is an i.i.d random variable:
\begingroup\makeatletter\def\f@size{7}\check@mathfonts
\begin{equation}\label{Eq_Derivation_2}
\varphi_{\boldsymbol{I}_{\text{PPP}}}(\omega) =  \exp\left(   - \mathbb{E}_{\boldsymbol{g}} \int_{\delta_o}^\infty \left[1-\exp\left(\jmath\omega \boldsymbol{g} \gamma_1 P_o (L_n^2+u^2)^{-\frac{\alpha}{2}}\right)\right] \lambda_I \mathrm{d}u  \right),
\end{equation}
\endgroup
representing the exact CF of multi-lane scenario, this integral can be evaluated numerically to obtain the cumulative distribution function (CDF) utilizing the Gil-Pelaez's inversion theorem \cite{Inversion_Theorem},
\begin{equation}\label{Eq_Inversion_Theorem}
F_{\boldsymbol{I}_{\text{PPP}} } (x) = \frac{1}{2} - \frac{1} {\pi} \int_0^\infty \frac{1} {\omega} \mathrm{Im}  \left[ \varphi_{\boldsymbol{I}_{\text{PPP}}}(\omega) \exp(-\jmath\omega x) \right] \mathrm{d}\omega.
\end{equation}
Further simplification can be achieved when the lane distance is ignored with respect to the longitudinal stretch of the road i.e. $L_n \to 0$, thus:
\begingroup\makeatletter\def\f@size{8.5}\check@mathfonts
\begin{align}\label{Eq_Derivation_3}
\varphi_{\boldsymbol{I}_{\text{PPP}}}(\omega)|_{L_n\to 0} = \lambda_I  &\delta_o - \mathbb{E}_{\boldsymbol{g}}\left[\lambda_I\frac{\delta_o}{\alpha} \mathcal{E}_{1+\frac{1}{\alpha }}\left(-\jmath \boldsymbol{g} \gamma_1 P_o \delta_o ^ {-\alpha } \omega \right)+ \right. \nonumber \\
& \left. \lambda_I\Gamma \left(1-\frac{1}{\alpha }\right) (-\jmath \boldsymbol{g} \gamma_1 P_o \omega )^{\frac{1}{\alpha } }\right],
\end{align}
\endgroup
where $\mathcal{E}_n(z)=\int_1^\infty\frac{e^{zt}}{t^n}\mathrm{d}t$ is the generalized exponential integral function.  However, in order to preserve the tractability of our approach, we further simplify the interference characteristic function by setting $\delta_o=0$, and taking the common case in radars where $\alpha\approx 2$, further by assuming a no channel fading, thus (\ref{Eq_Derivation_3}) becomes:
\begin{equation}\label{I_dist}
\varphi_{\boldsymbol{I}_{\text{PPP}}}(\omega)|_{\text{wc}}  = \exp \left(- \sqrt{-\jmath \pi \gamma_1 P_o \omega  \lambda_I^2}\right),
\end{equation}
having a tractable so-called L\'{e}vy distribution which can be seen as an inverse gamma distribution \cite{Haenggi_Book}, having a CF and a CDF of the form:
\begin{align}\label{Levy_dist}
\varphi(\omega) &= \exp \left(\jmath \mu \omega - \sqrt{-2 \jmath a \omega}\right), \nonumber \\
F_X(x) &= \mathrm{erfc}\left(\sqrt{\frac{a}{2(x-\mu)}}\right)
\end{align}
where $\mu$ is the location parameter, $a$ is the scale parameter, and $\mathrm{erfc}(z)$ is the complementary error function. By comparing (\ref{Levy_dist}) and (\ref{I_dist}) we can conclude that the interference follows a Levy distribution with CDF:
\begin{equation} \label{Eq_22}
F_I(x)|_{\text{wc}} = \mathrm{erfc}\left(\sqrt{\frac{\pi(\xi\lambda)^2 \gamma_1 P_o}{4x}}\right),
\end{equation}
and with location parameter $\mu=0$ and scaling parameter $a=\frac{1}{2}\pi(\xi\lambda)^2 \gamma_1 P_o$, noting here that the case of $\delta_o\to 0$ represents the worst-case scenario when antennas have very low directivity $\theta\approx 180^\circ$ leading to higher interference level. For brevity we call this scenario ($\delta\to 0,\alpha \approx 2$ and $L_n \to 0$) the \emph{worst case} where the subscript \emph{wc} is added. We depict in Fig. \ref{Fig_I_CDF_PPP} a plot of the interference CDF with PPP geometrical model. As it is expected, the assumption of \emph{no interference guard} $\delta_o=0$ leads to an increase in the interference level\footnote{Note that when the beamwidth $\theta$ gets wider the antenna gain $G_t$ gets lower, however, in order to establish a common ground for comparing \emph{wc} case with other cases, we assume that the product $\gamma_1 P_o$ is constant, thus $P_o$ is increased to compensate the reduction in $G_t$.}. The figure also shows consistency with Monte-Carlo simulations. The procedure of simulation is discussed in details in Sec. \ref{Sec_Simulation}.

\begin{figure}[!t]
	\centering
	\includegraphics[width=\linewidth]{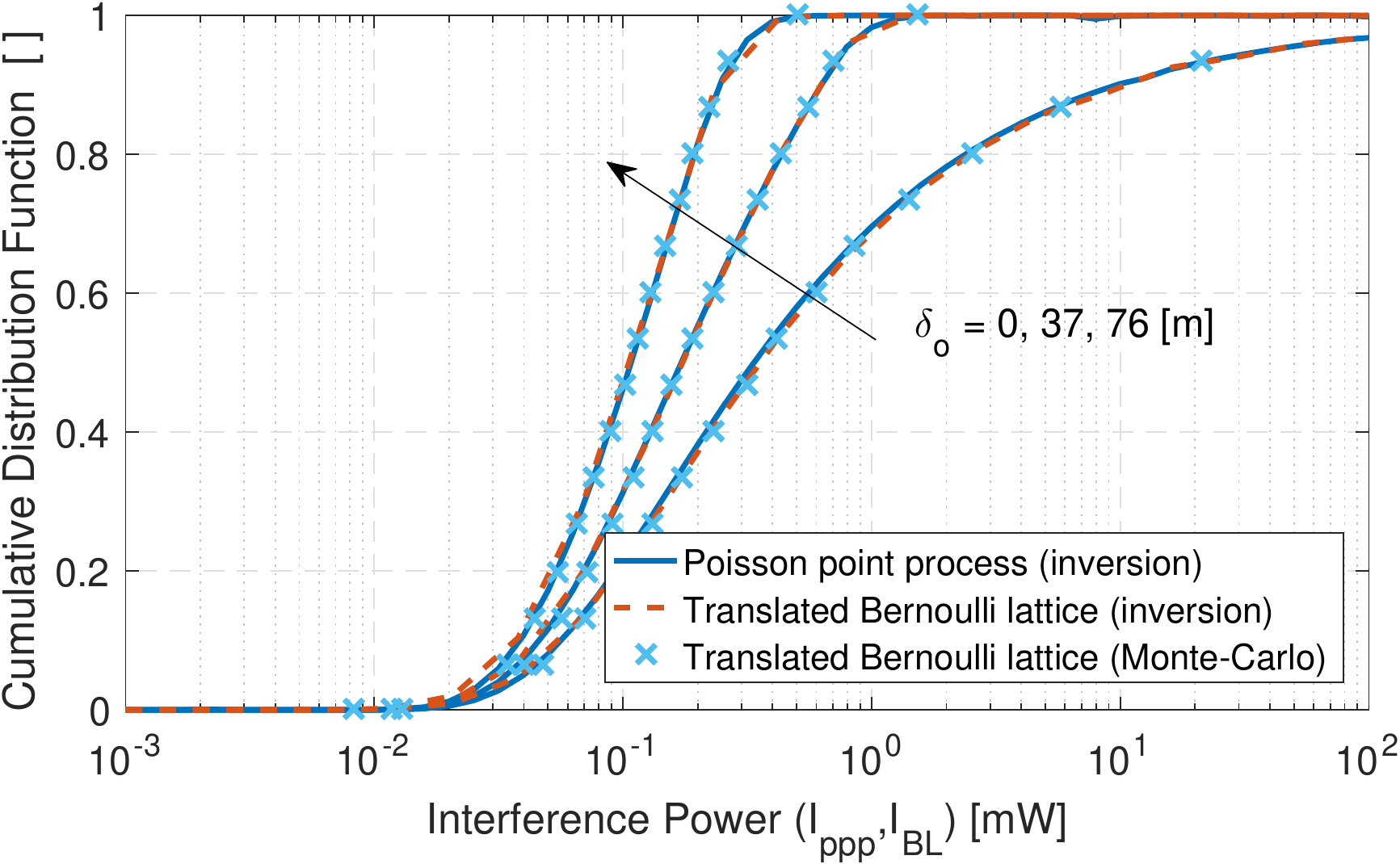}
	\caption{A comparison of CDF of the interference arising from the geometric PPP model and by the BL model, obtained using by the inversion theorem in (\ref{Eq_Inversion_Theorem}) and inverse Laplace transform in (\ref{Eq_Inv_Laplace}). Parameters as per Table \ref{Table_Notations}, and $\frac{1}{\lambda_I}=100$ m. }
	\label{Fig_I_CDF_BL}
\end{figure}

We utilize a similar approach to analyse the CDF of the interference caused by the translated Bernoulli lattice model. We start with the definition of Laplace transform of a random variable $\boldsymbol{X}$ as: $\mathcal{L}_s\{ \boldsymbol{X} \}=\mathbb{E} \left[ e^{-s \boldsymbol{X}} \right]$. Accordingly we can write:
\begingroup\makeatletter\def\f@size{6}\check@mathfonts
\begin{align}\label{Eq_Laplace_BL}
	&\mathcal{L}_s\{\boldsymbol{I}_{\text{BL}}\} = \mathbb{E} \left[ e^{-s \boldsymbol{I}_{\text{BL}}} \right] \nonumber \\
	&=\mathbb{E}_{\boldsymbol{g},\boldsymbol{\mathcal{U}},\boldsymbol{B}} \left[ \exp\left( -s \sum_{m=0}^{\infty} \boldsymbol{g}_m \boldsymbol{B}(\xi) \gamma_1 P_o
	\left[L_n^2 + (\delta_o +(m+\boldsymbol{\mathcal{U}})\delta)^2\right]^{-\frac{\alpha}{2}}  	 \right) \right] \nonumber \\
&=\mathbb{E}_{\boldsymbol{g},\boldsymbol{\mathcal{U}},\boldsymbol{B}} \prod_{m=0}^{\infty} \left[ \exp\left( -s  \boldsymbol{g}_m \boldsymbol{B}(\xi) \gamma_1 P_o \left[L_n^2 + (\delta_o +(m+\boldsymbol{\mathcal{U}})\delta)^2\right]^{-\frac{\alpha}{2}}  \right) \right] \nonumber \\
&\stackrel{(a)}{=}\mathbb{E}_{\boldsymbol{g},\boldsymbol{\mathcal{U}}} \prod_{m=0}^{\infty} \left[ (1-\xi)+\xi\exp\left( -s  \boldsymbol{g}_m  \gamma_1 P_o \left[L_n^2 + (\delta_o +(m+\boldsymbol{\mathcal{U}})\delta)^2\right]^{-\frac{\alpha}{2}}  \right) \right],
\end{align}
\endgroup
where final step (a) follows from applying the expectation over the discrete Bernoulli random variable $\boldsymbol{B}(\xi)$ noting that this RV is independent of the geometrical lattice. Utilising (\ref{Eq_Laplace_BL}) we can obtain the CDF of the interference by inverting the Laplace transform of $\boldsymbol{I}_{\text{BL}}$,
\begin{equation} \label{Eq_Inv_Laplace}
F_{\boldsymbol{I}_{\text{BL}}}(x)=\mathcal{L}^{-1}\left[\frac{1}{s}\mathcal{L}_s\{\boldsymbol{I}_{\text{BL}}\}\right],
\end{equation}
We utilize the Talbot inversion method with the unified numerical inversion frame work of \cite{Ward_Whitt_Inversion} in order to generate the plot in Fig. \ref{Fig_I_CDF_BL}. It is interesting to note that both the PPP model and the BL model share very similar CDFs, validated by running a large number of Monte-Carlo simulations. %The simulation parameters are $P_o=10$ dBm, $\xi=\frac{1}{10}$,  $\delta_o=\{0,37,76\}$ m, and a fixed interferers inverse-density of $\frac{1}{\lambda_I}=100$ m.

\subsection{The Convergence of Lattice to Poisson}
It is not surprising to anticipate the convergence of a Bernoulli lattice point process to a Poisson point process when the spacing of the lattice points decreases while reducing the thinning probability to hold a constant resulting intensity. This occurs because a repeated random translation of a point process increases its entropy \cite{Daley} and possibly leads to a Poisson point process that has maximum entropy. A Bernoulli lattice process $\Phi_{\mathrm{BL}}$ thinned out of a lattice $\Phi_L=\{ \delta \boldsymbol{x} ~ : ~ \boldsymbol{x} \in \mathbb{Z}^d\}$ (in the space $\mathbb{R}^d$) with a decreasing retention probability $\xi = \delta^d \lambda$, coverages in distribution to a Poisson point process of intensity $\lambda$ as $\delta \to 0 $. The intuition can be explained when we take an arbitrary family of bounded and disjoint Borel subsets of $\mathbb{R}^d$ denoted by $A_1,\dots,A_m \subset \mathbb{R}^d ~ , ~ \forall m \in \mathbb{N}^+$, based on which we define a modified set  $\tilde{A_1},\dots \tilde{A}_m \subset \mathbb{R}^d ~ , ~ \forall m \in \mathbb{N}^+$ constituting the union of the underlying lattice Voronoi cells that have their seeds included in the original Borel subsets, thus,
	\begin{equation}
	\tilde{A}_m = \bigcup_{x \in \{ \Phi_L \cap A_m\} } V(x),
	\end{equation}
	where $V(x)$ is the Voronoi cell of the seed $x$, defined as,
	\begin{equation} \label{Eq_Voronoi_Cells}
	V(x) \stackrel{\bigtriangleup}{=} \{u \in \mathbb{R}^d : ||x-u|| \leq ||x_i-u|| \, \forall x_i \in \Phi_L \setminus \{x\} \},
	\end{equation}
	Defining $\tilde{A}_m$ implicitly gives knowledge about its count of points included in the region (the counting measure on the lattice process) $\mathcal{N}_{\Phi_L}(\tilde{A}_m)=N_m$, where the number $N_m$ is $N_m = |\tilde{A}_m| / \delta^d = \lambda |\tilde{A}_m|$, where $|.|$ is the Lebesgue measure. Accordingly, we can write that the counting measure on the thinned lattice is an independent Binomial random variables with the following parameters.
	\begin{equation}\resizebox{.9\hsize}{!}{$
		\left\{\mathcal{N}_{\Phi_\mathrm{BL}}(\tilde{A}_1),\mathcal{N}_{\Phi_\mathrm{BL}}(\tilde{A}_2),\dots,\mathcal{N}_{\Phi_\mathrm{BL}}(\tilde{A}_m) \right\} \sim  \boldsymbol{\mathrm{Bin}}(\lambda |\tilde{A}_m|,\xi=\delta^d \lambda),
		$}
	\end{equation}
	and since a Binomial process coverages to a Poisson process as the number of trials goes to infinity, thus,
	\begin{equation}\label{Eq_Th_1}\resizebox{.9\hsize}{!}{$
		\left\{\mathcal{N}_{\Phi_\mathrm{BL}}(\tilde{A}_1),\mathcal{N}_{\Phi_\mathrm{BL}}(\tilde{A}_2),\dots,\mathcal{N}_{\Phi_\mathrm{BL}}(\tilde{A}_m) \right\} \underset{\delta \to 0}{   {\overset{d}{\rightarrow}} } \boldsymbol{\mathrm{Pois}}(\lambda),
		$}
	\end{equation}
	where $\overset{d}{\rightarrow}$ means converges in \emph{distribution}. In addition to the fact that $\tilde{A}_m$ is a pixelated version of $A_m$ and converges to $A_m$ as $\delta \to 0$,
	\begin{equation}\label{Eq_Th_2}
		\left\{\tilde{A}_1, \tilde{A}_2, \dots, \tilde{A}_m \right\} \underset{\delta \to 0}{\to} \left\{A_1,A_2, \dots, A_m \right\}
	\end{equation}
	We can conclude that the Bernoulli thinned lattice coverages to a Poisson point process since it satisfies the following two conditions for disjoint Borel subsets \cite{Book_Baddeley}:
	\begin{description}
		\item[Condition 1] The distribution of the count measures $\mathcal{N}_{\Phi_\mathrm{BL}}(A_1),\mathcal{N}_{\Phi_\mathrm{BL}}(A_2),\dots,\mathcal{N}_{\Phi_\mathrm{BL}}(A_m)$ are Poisson distributed by (\ref{Eq_Th_1}) and (\ref{Eq_Th_2}).
		\item[Condition 2] The count measures are independent, which follows directly from the assumption of the disjoint subsets and the independent thinning.
	\end{description}

\begin{figure}[!t]
	\centering
	\includegraphics[width=\linewidth]{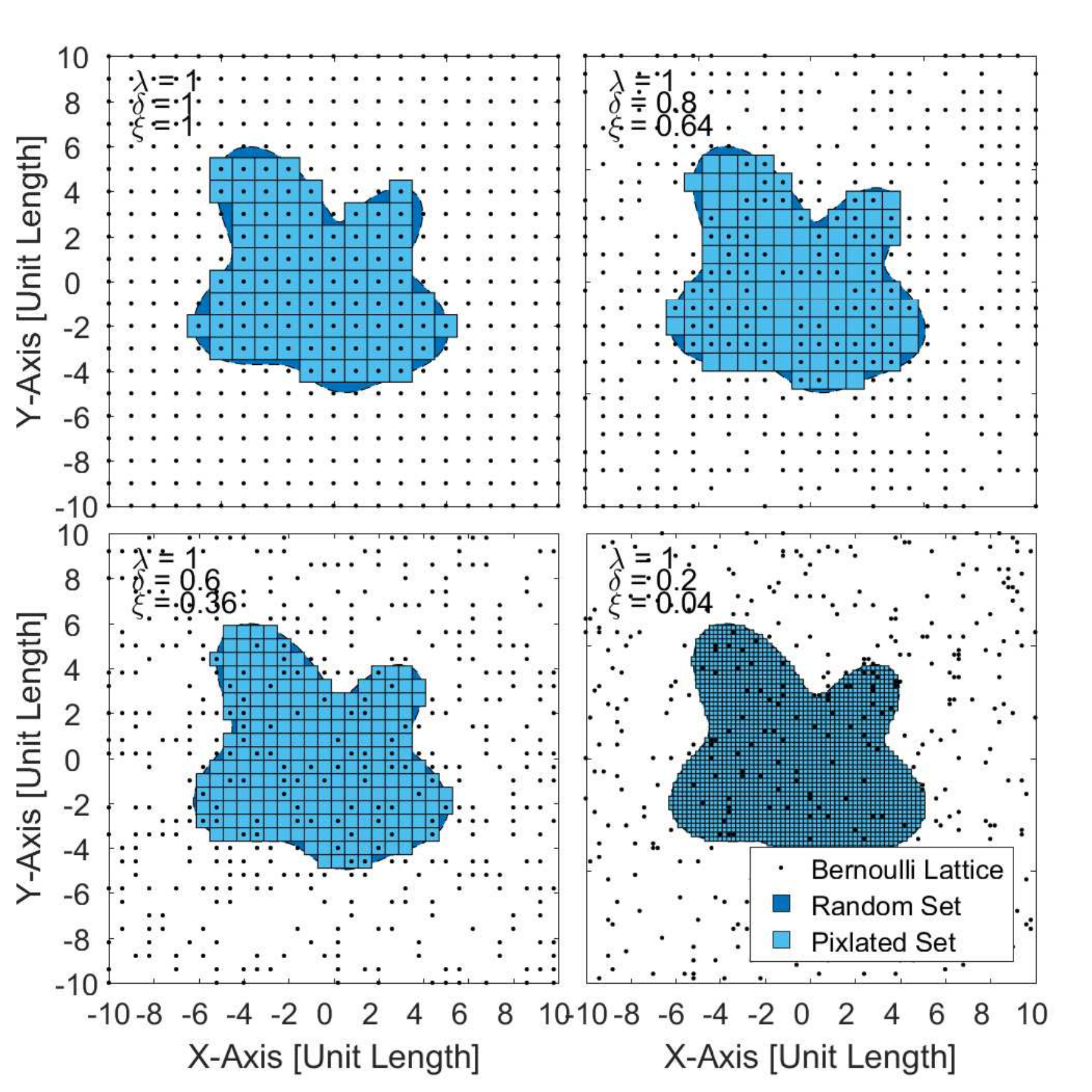}
	\caption{An illustration of the convergence of a Bernoulli lattice in two dimensional space to a Poisson point process as the lattice spacing $\delta$ vanishes, however preserving a constant points intensity of $\lambda=\frac{\xi}{\delta^2}$.}
	\label{Fig_Convergance}
\end{figure}

To illustrate the convergence of a Bernoulli lattice to a Poisson point process we depict in Fig. \ref{Fig_Convergance} a two-dimensional ($d=2$) representation of the thinned lattice and a random closed set $A_m$ with its pixelated version $\tilde{A}_m$.

\section{Radar Performance} \label{Sec_Performance}
If the aggregated interference $I$ summing at the receiver is uncorrelated with the transmit signal, then the receiver perceives this interference as noise-like, thus causing an increase in the noise floor. This behaviour is analytically investigated for FMCW radars in \cite{4106078} as well as emphasized by ITU-R in \cite{ITU_AutomotiveRadar}.  The performance of radar is thus limited by the signal to interference and noise ratio, defined as:
\begin{equation}\label{Eq_SIR}
\mathrm{SINR} = \frac{S}{\boldsymbol{I}+\boldsymbol{N}},
\end{equation}
where $\boldsymbol{N}$ is the noise power process generated in the receiver electronics, and $S$ is the reflected power from the target having the form described previously in (\ref{Eq_Modified_Radar_Equation}).

Having the $\mathrm{SINR}$ above a certain threshold $T$ leads to successful ranging and detection. Accordingly we form the probability of successful ranging as:
%\begin{align}\label{Eq_Ps}
%p_s  &=\mathbb{P}[\mathrm{SINR} \geq T | N] \nonumber \\
%     &=\mathbb{P}[\boldsymbol{I} \le \frac{S}{T}-N | N] = E_N \left[F_{\boldsymbol{I}}\left( \frac{S}{T}-N\right)\right],
%\end{align}

\begin{align}\label{Eq_Ps}
p_s  &=\mathbb{P}[\mathrm{SINR} \geq T ] \nonumber \\
     &=\mathbb{P}[\boldsymbol{I} \le \frac{S}{T}-N ] = F_{\boldsymbol{I}} \left(\frac{S}{T}-N\right),
\end{align}
%where $E_N[z]$ is the expectation over the noise random variable.
that can be easily calculated from (\ref{Eq_Inversion_Theorem}) or (\ref{Eq_Inv_Laplace}).

\begin{figure}[!t]
	\centering
	\includegraphics[width=\linewidth]{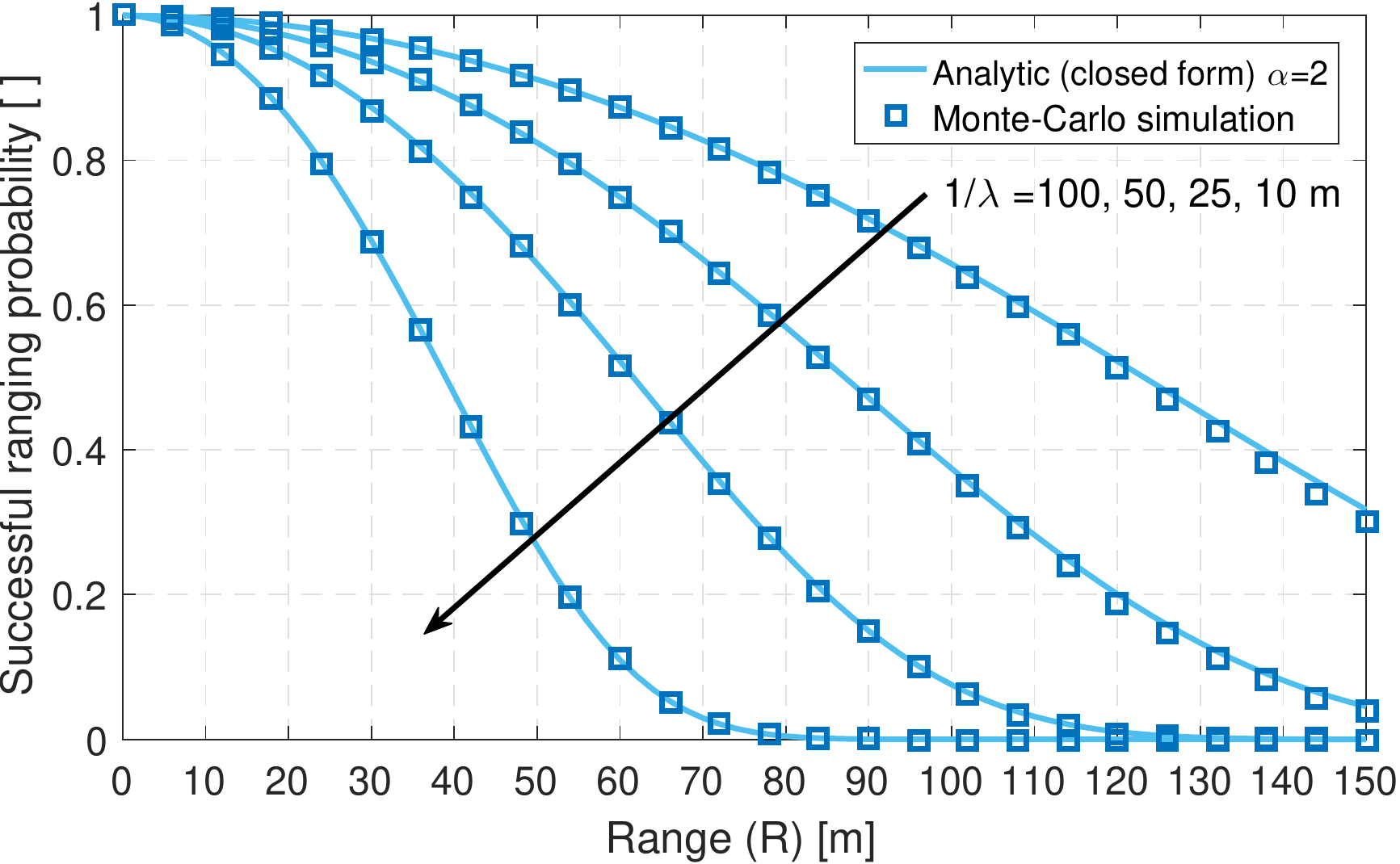}
	\caption{The ranging success probability $p_s$ in (\ref{Eq_Ps_main}) versus the ranging distance for the worst case scenario, using $\xi = \frac{1}{100}$. }
	\label{Fig_Proc_Success}
\end{figure}

\begin{figure}[!t]
	\centering
	\includegraphics[width=\linewidth]{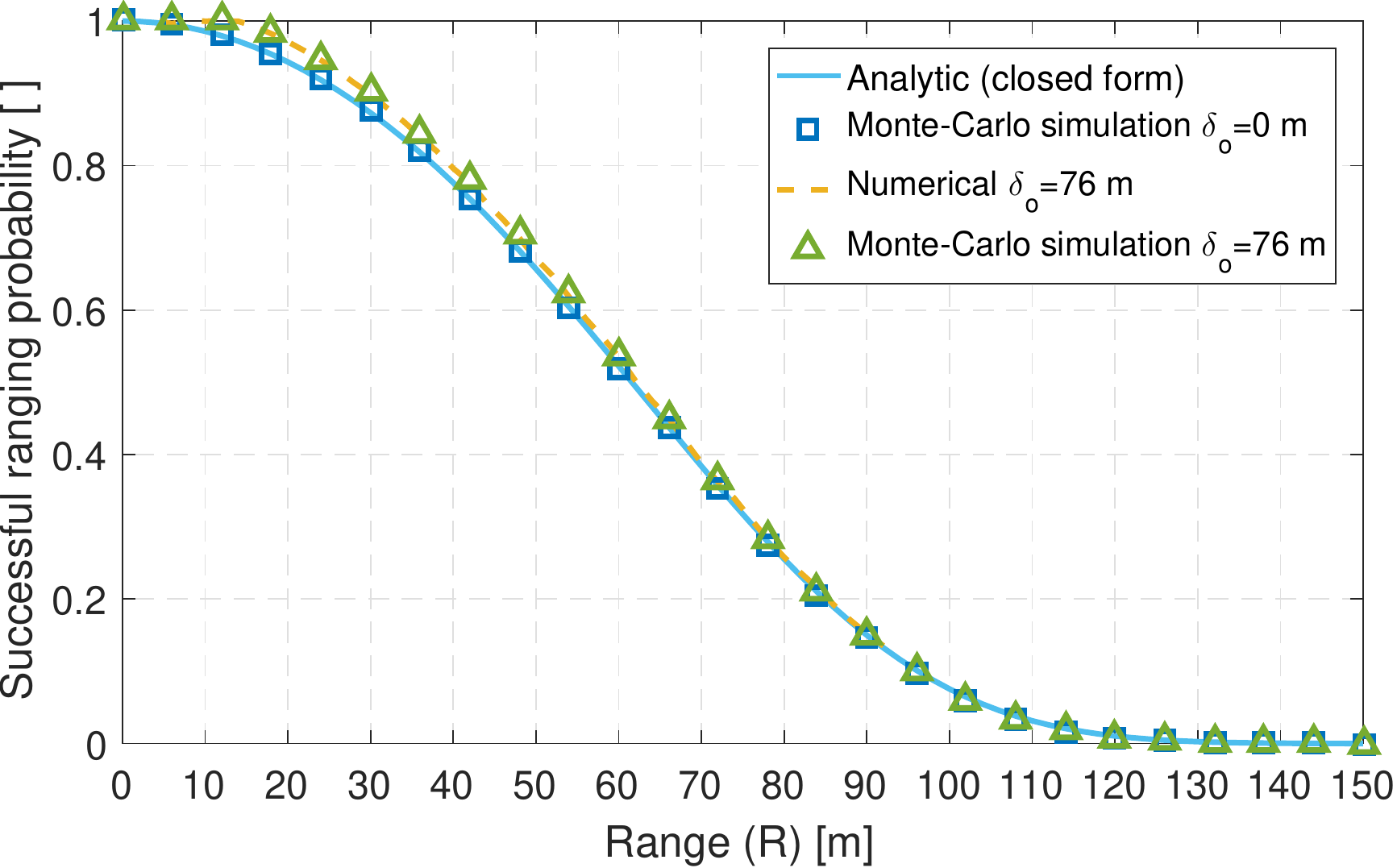}
	\caption{The ranging success probability $p_s$ in (\ref{Eq_Ps}) versus the ranging distance for PPP model using $\alpha=2$ and comparing the case of $\theta=15^\circ$ (i.e. $\delta_o=76$ m) and the tractable lower bound in (\ref{Eq_Ps_main}). Using interferers density $\lambda = \frac{1}{25}$ m, and $\xi = \frac{1}{100}$.}
	\label{Fig_ps_compare}
\end{figure}

As discussed previously in the worst case scenario the guard distance $\delta_o$ vanishes due the lack of antenna directivity, thus in this scenario the interference take a simple closed form (\ref{Eq_22}), and the radar ranging success probability becomes,
\begin{equation}\label{Eq_Ps_General}
p_s|_{\text{wc}} = \mathrm{erfc} \left(\sqrt{ \frac{\frac{\pi}{4}(\xi\lambda)^2 \gamma_1 P_o}{\frac{\gamma_1 \gamma_2 P_o R^{-4}}{T}+N} }\right),
\end{equation}
where this form is valid for the case $\alpha \approx 2$. For dense traffic conditions and sufficient radar power, the limiting performance factor becomes the interference rather than the noise. In this case, the ranging success probability will take a simpler form,
\begin{equation}\label{Eq_Ps_main}
p_s|_{\text{wc},N\to 0} = \mathrm{erfc} \left(\sqrt{\frac{\pi T}{4\gamma_2}}\xi\lambda R^2\right),
\end{equation}
This provides an insight on the main dynamics affecting the performance of automotive radar. In interference-limited environment, the ranging success probability is independent of the common transmit power $P_o$ and the antenna characteristics $\gamma_1$. We plot in Fig. \ref{Fig_Proc_Success} the ranging probability as a function of the ranging distance $R$, and the vehicles intensity $\lambda$ using the closed form in (\ref{Eq_Ps_main}) and compare with Monte-Carlo simulations.  We also plot in Fig. \ref{Fig_ps_compare} the ranging success probability for the general case as calculated from (\ref{Eq_Ps}) compared with the performance of the tractable scenario representing the lower performance bound in (\ref{Eq_Ps_main}).

\subsection{Radar Performance Optimization}
As it can be deduced from (\ref{Eq_Ps_main}) a higher spectrum collision probability $\xi$ leads naturally to a lower radar success rate. However by reducing $\xi$, the spectrum utilization efficiency will proportionally drop, since fewer vehicles are concurrently accessing the spectrum. If vehicles utilise a random spectrum access policy then $\xi$ can also be seen as the transmission probability over the shared bandwidth. Finding an optimum design value of $\xi$ can substantially enhance the overall system performance for all users.

\begin{figure}[!t]
	\centering
	\includegraphics[width=\linewidth]{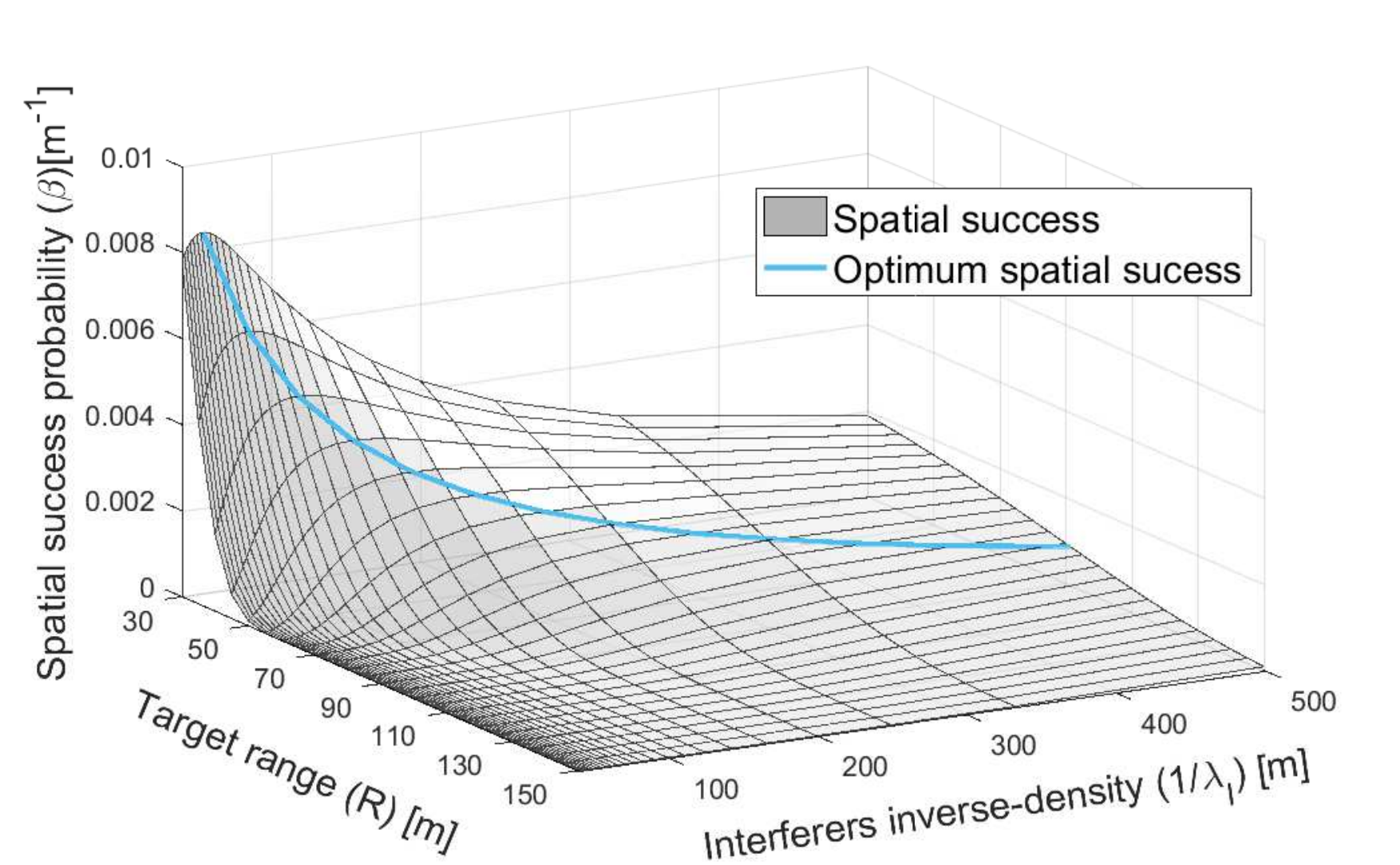}
	\caption{The spatial success probability $\beta$ given in (\ref{Eq_Spatial}) versus the ranging distance $R$ and the interferes density $\lambda_I$, showing the corresponding optima. }
	\label{Fig_Optimum}
\end{figure}

We define the \emph{spatial success probability} $\beta$ as the probability of successful spectrum access per unit length expressed as,
\begin{equation}\label{Eq_Spatial}
\beta = \lambda \xi p_s,
\end{equation}
representing the density of vehicles that are detecting their targets successfully. Recalling that $\lambda_I=\lambda \xi$, we plot in Fig. \ref{Fig_Optimum} the spatial success probability against $\lambda_I$, and the target range $R$, noting that a certain optimum density point exists for a certain target range, where operating at this point leads to maximizing the spatial success probability. As $\beta$ is a concave function, we formulate the optimum intensity as,
\begin{equation} \label{Eq_Lambda_Opt_1}
\lambda_{I}^{*} = \underset{\lambda_I}{\text{argmax}} \left[ \beta = \lambda_I \text{erfc}(C \lambda_I) \right], \end{equation}
taking $\frac{\partial \beta}{\partial \lambda_I}=0$ yields,
\begin{equation}\label{Eq_Lambda_Opt_2}
 \text{erfc}(C \lambda_I^{*})=\frac{2 C \lambda_I^{*} e^{-C^2 {\lambda_I^{*}}^2}}{\sqrt{\pi }} 
\end{equation}
where $C=\sqrt{\frac{\pi T}{4\gamma_2}} R^2$. However, no exact closed form solution for (\ref{Eq_Lambda_Opt_2}) exists, alternatively we define a new variable $z_o=C\lambda_I^{*}$ and substitute in (\ref{Eq_Lambda_Opt_2}) leading to a numerical solution of $z_o \approx 0.532$, accordingly the optimum transmission probability is:
\begin{align}\label{Eq_Optimum_xi}
\xi^{*}=\min\left[\frac{z_o}{\lambda C},1\right]=\min\left[\frac{z_o}{\lambda}\sqrt{\frac{4\gamma_2}{\pi T}}R^{-2},1\right],
\end{align}
where the function $\min\left[.,.\right]$ ensures that the transmission probability is less than unity. Accordingly, the optimum spatial success probability is given by substituting in (\ref{Eq_Spatial}),
\begin{align}
\beta^* &= \lambda \xi^* \mathrm{erfc} \left(C\lambda \xi^*\right) \nonumber \\
&=\lambda \min\left[\frac{z_o}{\lambda C},1\right] \mathrm{erfc} \left(\lambda C \min\left[\frac{z_o}{\lambda C},1\right]\right),
\end{align}

\begin{figure}[!t]
	\centering
	\includegraphics[width=\linewidth]{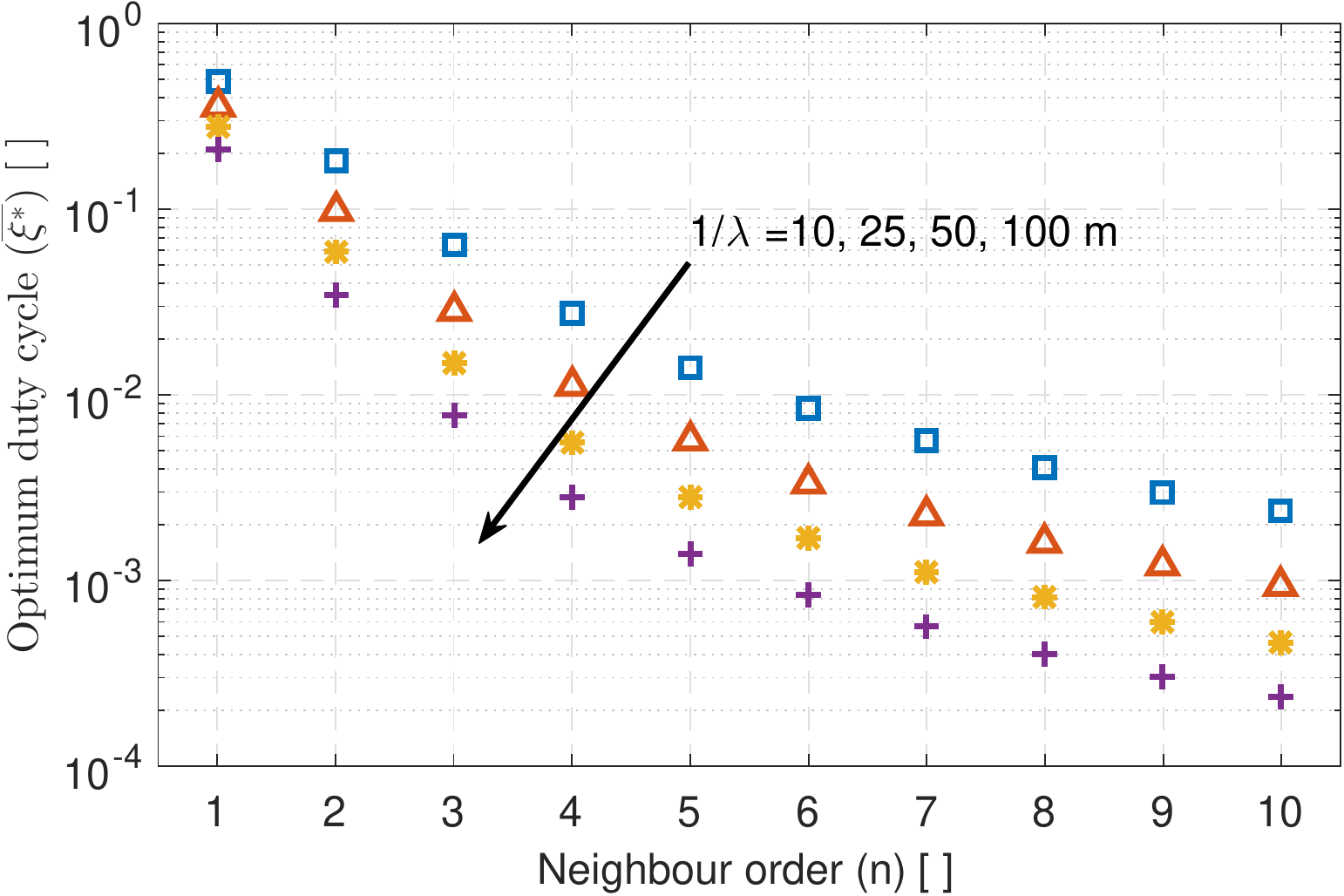}
	\caption{Optimum duty cycle $\overline{\xi^*}$ given in (\ref{Eq_xi_vs_n}) versus the design value of the target $n^{th}$ neighbour, for different vehicle densities.}
	\label{Fig_xi_opt}
\end{figure}

However, the ranging distance $R$ is a stochastic quantity with statistics depending on the vehicles linear density $\lambda$, where the $n^{\text{th}}$ nearest vehicle has a known closed-form distribution in a PPP process of,
\begin{equation}
f_{R_n}(r) = \frac{e^{-\lambda r } (\lambda  r)^n}{r \Gamma (n)} ~,~ \forall n \in \mathbb{N}^+,
\end{equation}
this follows from Eq. (2.21) in \cite{Haenggi_Book}, where $ n \in \mathbb{N}^+$ represents the order of the nearest vehicle within the same lane. We can optimise the transmission probability (or duty cycle) by setting a certain target number of nearest vehicles to be detected, so the average optimum is obtained by performing a statistical expectation over the contact distance $R$,
\begin{align}\label{Eq_xi_vs_n}
\overline{\xi^*}&= \mathbb{E}_R \left[\xi^*\right] \nonumber \\
&=\int_{0}^{\infty} \min\left[\frac{z_o}{\lambda}\sqrt{\frac{4\gamma_2}{\pi T}}R^{-2},1\right] \times   \frac{e^{-\lambda r } (\lambda  r)^n}{r \Gamma (n)}  \mathrm{d}r \nonumber \\
&=\frac{\lambda K \Gamma \left(n-2,\sqrt{K \lambda} \right)-\Gamma \left(n,\sqrt{K \lambda}  \right)+\Gamma (n)}{\Gamma (n)},
\end{align}
%\begin{align}\label{Eq_xi_vs_n}
%\overline{\xi^*}[n]&= \mathbb{E}_R \left[\xi^*\right] 
%=\int_{0}^{\infty} \frac{z_o}{\lambda}\sqrt{\frac{4\gamma_2}{\pi T}}r^{-2} \times   %\frac{e^{-\lambda r } (\lambda  r)^n}{r \Gamma (n)}  \mathrm{d}r \nonumber \\
%&= \frac{\lambda K}{(n-1)(n-2)} ~~,~\forall n>2
%\end{align}
where $K$ is a constant given by $K=z_o\sqrt{\frac{4\gamma_2}{\pi T}}$, and $\Gamma(a,x)=\int_x^\infty t^{a-1} e^{-t} \mathrm{d}t$ is the incomplete gamma function. %Similarly, we can find the corresponding optimum spatial success probability,
%\begin{equation}
%\overline{\beta^*}[n]= \frac{\mathrm{erfc}(z_o) \lambda^2 K}{(n-1)(n-2)}, ~~,~\forall n>2.
%\end{equation}
Using different vehicle densities and parameter values in Table \ref{Table_Notations}, we plot in Fig. \ref{Fig_xi_opt} the optimum transmission probability (or duty cycle) at which the a maximum spatial ranging success rate is achieved. As expected, the optimum transmission duty cycle drops with measured detection range, this becomes more clear if we study the asymptotic function of $\overline{\xi^*[n]}$ in (\ref{Eq_xi_vs_n}) as $n\to \infty$,
\begingroup\makeatletter\def\f@size{9.5}\check@mathfonts
\begin{align}\label{Eq_Asmptote}
\hat{\xi}[n] &= O\left(\overline{\xi^*[n]}\right)~\text{as}~ n\to \infty \nonumber \\
&\stackrel{(a)}{=} \underset{n\to \infty}{O} \left(1+  \frac{ \lambda K e^{-\sqrt{\lambda K}} e_n(\sqrt{\lambda K})}  {(n-1)(n-2) } -e^{\sqrt{\lambda K}} e_n(-\sqrt{\lambda K})\right) \nonumber \\
&=\frac{\lambda K}{n^2},
\end{align}
\endgroup
where step (a) follows from $\Gamma(n,x) = (n-1)!e^{-x} e_n(x)$ for $n\in\mathbb{N}^+$, and $e_n(x) = 1+x+x^2/2!+\dots+x^n/n!$ is the partial sum of the exponential series. Thus, we can clearly note from (\ref{Eq_Asmptote}) that detecting farther vehicles (higher $n$) will decrease the spectral efficiency, however it is interesting to note that an increased vehicle density leads to a better spectrum utilization, this is also apparent in Fig. \ref{Fig_xi_opt} comparing the $\xi$ for different values of $\lambda$.

\section{Simulation Procedure} \label{Sec_Simulation}
As it was illustrated previously in Figs. \ref{Fig_I_Mean}, \ref{Fig_I_CDF_PPP}, \ref{Fig_I_CDF_BL}, \ref{Fig_ps_compare}, and \ref{Fig_Proc_Success}, Monte-Carlo simulations showed a very close match with the analytical formulae. Firstly the vehicles are deployed over a length of 10,000 m, and the interference is summed at the origin as per (\ref{Eq_8}) and stored. The scenario is repeated at least 5,000 times and the statistics of the interference are obtained for Figs. \ref{Fig_I_Mean}, \ref{Fig_I_CDF_PPP}, and \ref{Fig_I_CDF_BL}. While for Fig. \ref{Fig_ps_compare} and \ref{Fig_Proc_Success} we further calculate the SINR as per (\ref{Eq_SIR}) and count the number of times it exceeds the threshold $T$, accordingly the simulated success probability is obtained from the following ratio:
\begin{equation*}
\hat{p}_s = \frac{\sum_{m=1}^{N} \mathbbm{1}_{\text{SINR}\ge T}}{N},
\end{equation*}
where $N$ is the total number of simulation runs. Then the process is repeated for all ranging $R$ sampling points.

\begin{table}[!htbp]
	\caption{Notations and Symbols}
	\centering
	\begin{tabularx}{\linewidth}{l l l}
		\hline\hline
		Symbol                   & Numerical Value                         & Explanation                     \\ \hline
		$R$                        & variable [m]                            & Distance to target              \\
		[0.5ex]
		$S$                & variable [W]                            & Reflected (bounced) signal      \\
		[0.5ex]
		$L_n$              & 10 [m]                                  & Lane spacing                    \\
		[0.5ex]
		$\delta_o$         & Calculated from (\ref{Eq_delta_o})  [m] & Min. interference distance   \\
		[0.5ex]
		$\delta$           & $1/\lambda$ [m]                         & vehicles uniform spacing            \\
		[0.5ex]
		$P_o$              & 10 dBm \cite{ITU_AutomotiveRadar}       & Transmit power to antenna  \\
		[0.5ex]
		$\alpha$           & Variable                                & Path-loss exponent              \\
		[0.5ex]
		$f$                & 76.5 GHz  \cite{ITU_AutomotiveRadar}    & Centre frequency               \\
		[0.5ex]
		$G_t$              & 45 dBi  \cite{ITU_AutomotiveRadar}      & Max. antenna gain               \\
		[0.5ex]
		%$A_e$             & 1  m$^2$                                & Effective area of the antenna   \\
		$\theta$           & 15$^\circ$ \cite{ACMA_Radar}            & Antenna beamwidth               \\
		[0.5ex]
		$\sigma_c$         & 30 dBsm \cite{6127923}                  & Radar cross-section             \\
		[0.5ex]
		$T$                & 10 dB \cite{6127923}                    & SIR threshold                   \\
		[0.5ex]
		$\boldsymbol{g_x}$ & -                                       & Interfering signals fading processes   \\
		[0.5ex]
		$\lambda$          & -                                       & Vehicles linear intensity       \\
		[0.5ex]
		$\lambda_I$        & -                                       & Interferers effective intensity \\
		[0.5ex]
		$\xi$              & 0.1                                     & Spectrum collision probability  \\
		[0.5ex]
		$\xi^*$              & refer to (\ref{Eq_Optimum_xi})                                   & Optimum duty cycle  \\
			[0.5ex]
		$z_o$              & 0.531597                                & Constant                        \\[0.5ex]
		$\beta$    & refer to (\ref{Eq_Spatial}) & Spatial success probability \\
		[0.5ex]
		$c$                & $\approx$ 3E8  [m/s]                    & Speed of light                  \\
		[0.5ex]
		$\gamma_1$         &  refer to (\ref{Eq_Gammas})                       & Radar-specific constant         \\
		[0.5ex]
		$\gamma_2$         &   refer to (\ref{Eq_Gammas})                     & Target-specific constant         \\
		[0.5ex]
		$C$         &   $\sqrt{\frac{\pi T}{4\gamma_2}} R^2$                     & -        \\[0.5ex]
		$K$			&	$z_o\sqrt{\frac{4\gamma_2}{\pi T}}$ & - \\			[0.5ex]
		\hline             &  \\ 
		[1.0ex]                    &
	\end{tabularx}
	\label{Table_Notations}
\end{table}

\section{Conclusion} \label{Sec_Conclution}
In this paper we introduced a novel approach in modelling automotive radar interference based on stochastic geometry for two point models. The first model assumes no correlation between vehicle locations, namely using a Poisson point process, while the second model assumes the vehicle locations on a fully regular lattice. We developed a framework to analytically calculate the mean value of the interference as well as its cumulative distribution function. We showed that under certain circumstances the cumulative distribution function can be reduced to explicit expression. Results suggested that both Poisson and lattice models cause very similar interference statistics under practical system parameters. In order to understand the similarity between the Poisson model and the Bernoulli lattice model, we presented an intuitive explanation of their convergence.

We further utilized the developed model to estimate the success probability of automotive radars in detecting their targets. This probability is based on the vehicle density, transmit power, antenna beamwidth, lane separation distance and most importantly on the duty cycle of the random spectrum access.
An optimization methodology is presented to design the optimum random spectrum access duty cycle and a closed form expression is provided for the optimum value. Future work will include the time evolution of the interference taking into consideration a non stationary traffic models.

\section*{Acknowledgement}
This work was supported by the Australian Research Council Discovery Project Grant ``Cognitive Radars for Automobiles'', ARC grant number DP150104473.
\ifCLASSOPTIONcaptionsoff
  \newpage
\fi

\bibliography{Radar_Interference}

\end{document}